\documentstyle[11pt, epsf]{article} 
\normalbaselines

\newcommand{\be}{\begin{equation}}
\newcommand{\ee}{\end{equation}}

\begin{document}

\begin{center}
{\bf To appear in {\it Journal of Integrative Neuroscience} 11 (2), June 2012.}

\vspace{4mm}
{\Large \bf{A Quantum-mechanical description of ion motion within the confining potentials of voltage gated ion channels}}

\vspace{3mm}
Johann Summhammer$^{a}$, Vahid Salari$^{b}$, Gustav Bernroider$^{c,}${\footnote{corresponding author}

\vspace{3mm}
{$^{a}$ Atom Institute, Vienna University of Technology, \\
Stadionallee 2, A-1020 Vienna, Austria\\
}

{$^{b}$ Kerman Neuroscience Research Center (KNRC),\\ Kerman, Iran
}

{$^{c}$ Department of Organismic Biology, Neurosignaling Unit, University of Salzburg\\ 
Hellbrunnerstr 34,A-5020 Salzburg, Austria\\
Email: gustav.bernroider@sbg.ac.at} 
}
\end{center}

\vspace{5mm}

{\bf Abstract}

{\small
Voltage gated channel proteins cooperate in the transmission of membrane potentials between nerve cells.  With the recent progress in atomic-scaled biological chemistry it has now become established that these channel proteins provide highly correlated atomic environments that may maintain electronic coherences even at warm temperatures.  Here we demonstrate solutions of the Schr\"{o}dinger equation that represent the interaction of a single potassium ion within the surrounding carbonyl dipoles in the Berneche-Roux model of the bacterial \textit{KcsA} model channel. We show that, depending on the surrounding carbonyl derived potentials, alkali ions can become highly delocalized in the filter region of proteins at warm temperatures.  We provide estimations about the temporal evolution of the kinetic energy of ions depending on their interaction with other ions, their location within the oxygen cage of the proteins filter region and depending on different oscillation frequencies of the surrounding carbonyl groups. Our results provide the first evidence that quantum mechanical properties are needed to explain a fundamental biological property such as ion-selectivity in trans-membrane ion-currents and the effect on gating kinetics and shaping of classical conductances in electrically excitable cells.

\noindent
Keywords: Quantum mechanics; quantum biology; quantum oscillations; Schr\"{o}dinger equation;  neural signaling; molecular dynamics;  ion channels; coherence dynamics; selectivity filter;  filter gating. 
}

\section{Introduction}

Ion channels are the building blocks of electrical membrane signals in the nervous system. They are responsible for the controlled charge transition across cell membrane, building up the trans-membrane potentials that propagate along the membranes as either dendritic potentials or action potentials. The 'channels' are proteins inserted into the cell membrane and host different sub-domains that operate at highly different physical action orders, ranging from the quantum-scale at the single atom level up to the classical scale of the entire protein \cite{GB1,GB2}. The main parts of the protein consist of the pore domain gate, controlling the access of ions into the protein, the cavity region that hosts a hydrated ion prior to its access to the narrow 'selectivity filter' that provides the last steps of charge transfer into the cell or outside the cell (see Figure 1). The selectivity filter (SF) of ion channel proteins is responsible for the selective and fast conduction of ions across neuronal cell membranes. After the determination of the atomic resolution structure of the bacterial model KcsA channel by MacKinnon and colleagues \cite{Mac,B1}, it became increasingly apparent that delicate atomic interactions within the highly ordered conduction pathway of the SF are critical for the ability of the protein to provide a very high rate of conduction without loss of selectivity for a particular ion species. Following these initial findings a long list of molecular dynamics studies (MD) added important functional aspects to the static details of the channel protein as reviewed by \cite{B2,corry,Ku,Roux}. Density functional calculations and hybrid QM/MM methods have substantially contributed in capturing the interaction between carbonyls and their effect on backbone atoms \cite{Bu}, together with a significant charge-polarization effect of ions on their filter ligands \cite{Guid}. Taking into account these results, additional theoretical treatments of ion complexation in water and diverse binding site models have revealed the topology of forces that are engaged in providing a highly selective coordination of ions in the filter region \cite{Bost}.
 
	The question became dominant how selectivity can emerge without compromising conduction.  Within this context at least two sets of problems are at the center of interest. First, it can be expected that the Coulomb type forces among ions , the adjacent water dipoles and the neighboring oxygen atoms, with average distances of 0,285 nm in the KcsA filter \cite{Kona}, require a quantum description, because the involved distances are near the de Broglie wavelength associated with the ions at thermal energies.  Are there typical quantum interaction effects or quantum interferences that are indispensible to explain the biological functioning of these channel proteins ?  The second point involves the question, can quantum effects propagate into the classical states of proteins ?  The last question gains considerable significance from findings about different 'gating mechanisms' of the channel. Whereas 'pore gating' is a relatively slow process that controls the access of ions to the cavity region of the filter, the filter itself is found to be able to change between 'permissive' and 'non-permissive' conformational states at a much shorter time-scales \cite{B3,Roux}. Because 'selectivity' seems to be organized within the filter that is located opposite the 'entrance' of the proteins pore region, it can be expected that the mechanisms providing selectivity must somehow become coordinated with the mechanisms controlling access of ions to the cavity region. If quantum effects do play a critical role for filter-ion coordination, it is feasible that these delicate interactions could leave their quantum traces in the overall conformation and the molecular gating state of the entire protein.
	
	In the present paper we investigate this question. We study the question  if and how quantum interferences in the filter region of a typical K-channel can influence the atomic environment within the frame set by time-dependent potentials derived from carbonyl charges, water molecules and other ions present in the atomic environment of the filter-region. By solving the Schr\"{o}dinger equation for the quantum mechanical states of $K^{+}$ ions, we demonstrate that, depending on the size of the confining potential and the thermal energy of the ion, the ion's wave-function can become highly delocalized, with its probability distribution extending over a significant fraction of the filter region. This in turn exerts an influence on the motion and induced polarization of the carbonyl groups lining the filter in a way that is different from the interaction that could be expected by a strictly  localized classical ion. Due to the interaction of the ions with a time-dependent potential, the energy of ions is not conserved. Instead, we observe tera-hertz (THz) oscillations of ionic wave-packets that become damped, giving off their energy to the environment either via the vibrational modes of the surrounding carbonyl dipoles or by radiation, or both. These effects will cool down the ions in the filter domain dramatically. We finally demonstrate the result of a 'replacement study', where $K^{+}$ ions in the same environment are replaced by Na+ ions. The results show that the oscillation frequency of the coordinating carbonyl dipoles discriminates the kinetic energy minima for both ion species. This is the first indication of a quantum mechanical effect on the ion selectivity function of voltage gated channel proteins.

\begin{figure}
\begin{center}
\epsffile{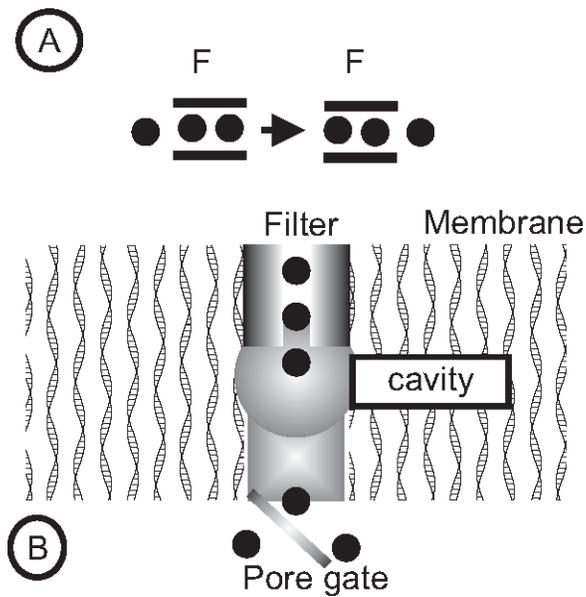}
\caption[Fig.1:]{The main physical operation facilitated by an ion channel-protein is an ion-species selective charge transfer (A), where one charge is taken up from the cavity (shown left), dehydrated and transferred into the filter region (marked by double bars). The filter region hosts at least two ions. The uptake of one ion leads to a knock-on effect on the ions present in the filter region. One ion is taken up from the cavity site, and another ion is released to the (intra or extra) cellular site (shown right). (B) The basic organization of voltage gated ion channels. The protein consists of three subdomains, the pore, the cavity and the filter region.}
\end{center}
\end{figure}

\section{Methods}

\subsection{The molecule}
We represent the selectivity filter as an array of C=O dipoles provided by the backbone carbonyl groups of the highly conserved TVGYG amino acid sequence (see Figure 1,2). Each filter segment carries five dipoles. We account for the intermittent, nearest neighbor water molecules in the filter by a confining potential for the ions. This water-derived potential behaves similar to repelling atoms, degrading the spread of the ion's wavepacket. The distances of carbonyl C-atoms to the filter axis were preset to 0,345 nm according to the Garafoli-Jordan permeation model \cite{Gara}. Effective charge distances of C=O dipoles were set to 0.122 nm. Bond directions along C=O groups are variable with zero degrees direction perpendicular to the filter axis. In the present hybrid model we consider one potassium ion as a quantum particle and treat the surrounding nuclear motion classically.

\begin{figure}
\begin{center}
\epsffile{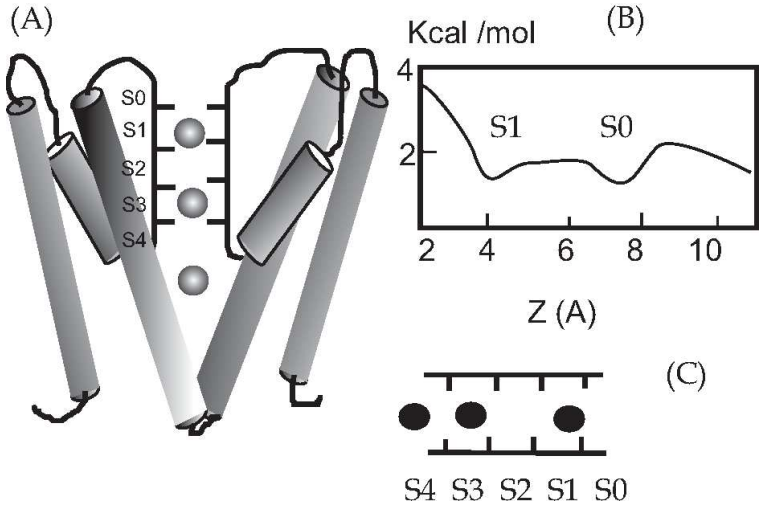}
\caption[Fig.2:]{A schematic illustration of a dimeric KcsA view with filter location and associated trans-membrane helices. The carbonyl atoms are indicated by short segments and numbered S0-S4 starting from the extracellular site. In (B)a demonstration of a free energy distribution given in kcal/mol along the axial z-coordinates (in Angstroems) of the filter according to the Berneche-Roux model is shown. The ion-filter configuration for the energy distribution in (B) is demonstrated by the scheme(C). In the present paper the ion at location S1 (insert C) is considered quantum-mechanically, the surrounding atoms from carbonyl groups, other ions in the filter region and the intermittent water molecule are treated as classical particles.}
\end{center}
\end{figure}

\subsection{The quantum model}

The basic interaction between classical ions and C=O dipoles as well as a quantum-mechanically described ion and it's atomic surrounding is the electrostatic repulsion and attraction. As usual the potential between two electrical point charges is

\begin{equation}
\ V(r)= \frac{1}{4\pi \epsilon}\cdot \frac{q_{1}q_{2}}r
\label{eq1}
\end{equation}

We have used $'r'$ for the distance between two particles and 'q' for their charges.  The symbol $'\epsilon'$ denotes the dielectric constant of the medium containing the charges. Taking into account a total of 20 C=O dipoles the resulting potential for the region containing one $K^{+}$ ion becomes:

\begin{equation}
\ V_{F}(\overline{r}, t) = \frac{q_{ion}}{4\pi\epsilon}  \sum^{20}_{j=1} \left(\frac{q_{O_{j}}}{\left|\overline{R}_{O_{j}}(t) - \overline{r}\right|} + \frac{q_{C_{j}}}{\left|\overline{R}_{C_{j}}(t) - \overline{r}\right|}\right)
\end{equation}

Here the vector $\bar{r}$ represents the ions position at time t, $q_{O_{j}}$ for the charge of the j-th oxygen and $q_{C_{j}}$for the charge of the j-th carbon atom respectively. Further $\bar{R}_{O_{j}}$ denotes the time-dependent position of the j-th oxygen and $\bar{R}_{C_{j}}$ the position of the j-th carbon atom. 

In additon we have to consider the potential V\textsc{i} that is contributed by the classical ions in the channel:

\begin{equation}
\ V_{I}(\overline{r}, t) = \frac{q_{ion}}{4\pi\epsilon}   \left(\frac{q_{K}}{\left|\overline{R}_{K_{1}}(t) - \overline{r}\right|} + \frac{q_{K}}{\left|\overline{R}_{K_{2}}(t) - \overline{r}\right|}\right)
\end{equation}

The potential due to the water molecules on both sides of the ion is given by a potential $V_{H_{2}O}$, which strongly depends on the form of the wave-packet. The time-dependent locations of classical ions have been denoted by $R_{K_{j}}$. The total potential to which the quantum particle is exposed is thus:\\
\begin{equation}
\ V(\bar{r}, t)= V_{F} (\bar{r},t) + V_{I} (\bar{r},t)+ V_{H_{2}O}(\bar{r},t)
\end{equation}

From a modelistic view the overall axial symmetry of the channel (Figure 3) allows for a restriction of the calculations to one spatial dimension, along the z-axis of the filter. This also reduces the computational effort considerably. Within the frame set we obtain a Schroedinger equation for one $K^{+}$ in the filter:\\
\begin{equation}
\ \frac{-\hbar^{2}}{2m_{k}} \frac{d^{2}}{dz^{2}} \psi(z,t) + V(z,t).\psi(z,t) = \textit{i}\hbar \frac{\partial}{\partial t} \psi (z,t)
\end{equation}
                                                                                   
The water molecules interacting with the ion move with the ion and thus somewhat impede the spreading of its wave-packet by an effective potential:

\begin{equation}
\ V_{H_{2}O}(z,t) = -E_{H_{2}O}.n\int^{d}_{-d}\left|\psi(z + s,t)\right|^{2}ds
\end{equation}

Here, $E_{H_{2}O}$  denotes the depth of the confining potential, set to 10 meV, which corresponds to about half the thermal energy $k_{B}T$.  The symbol 'd' represents  the average distance between the oxygen of a water molecule and the $K^{+}$ ion as available from both, experimental  \cite{Mae,Zhou} and theoretical studies \cite{Zhou}. For the present study we have set a d = 0.28 nm. Further 'n'  is a normalization constant which ensures that the minimum of $V_{H_{2}O}(z,t)$ equals  $-E_{H_{2}O}$.
We have calculated the wave-function  together with the resulting space and time dependent probability $p = \left|\psi(z,t)\right|^{2}$   for different oscillations of C=O groups and locations of classical ions in the filter region.  All second order differential equations were solved by applying the Crank-Nicolson formalism \cite{th}which is known for it's stability over a large range of calculational steps. Additional controls were applied to check the consistency of calculations by doubling the temporal resolution within a constant time interval (e.g. 678.500 steps with 0.006 fs to 2.750.000 steps with 0,0015 fs resolution, see Figure 4). No numerical artifacts on the stability of the results could be observed. Also, under constant, time independent potentials the wave-packets behavior was as can be expected for harmonic oscillating potential forms, for which analytic solutions are well known. The initial wave-function was approximated by a Gaussian distribution centered at one of the possible filter locations (S0-S4).  

\begin{figure}
\begin{center}
\epsffile{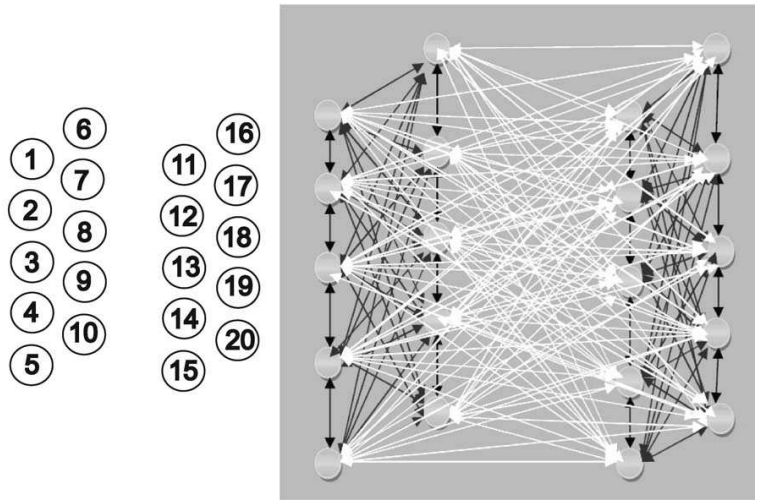}
\caption[Fig.3:]{A three-dimensional distance model for the carbonyl-ion coordination within the filter region of the KcsA model channel as used for the present study. The 20 C=O dipoles are symbolized by circles and are numbered on the left to specify the distance matrix that we used for the present calculations and the 3-dim topology provided by the right-hand figure (dark lines join the C=O dipoles from two lateral strands and white lines join opposing C=O dipoles)}
\end{center}
\end{figure}

\begin{figure}
\begin{center}
\epsffile{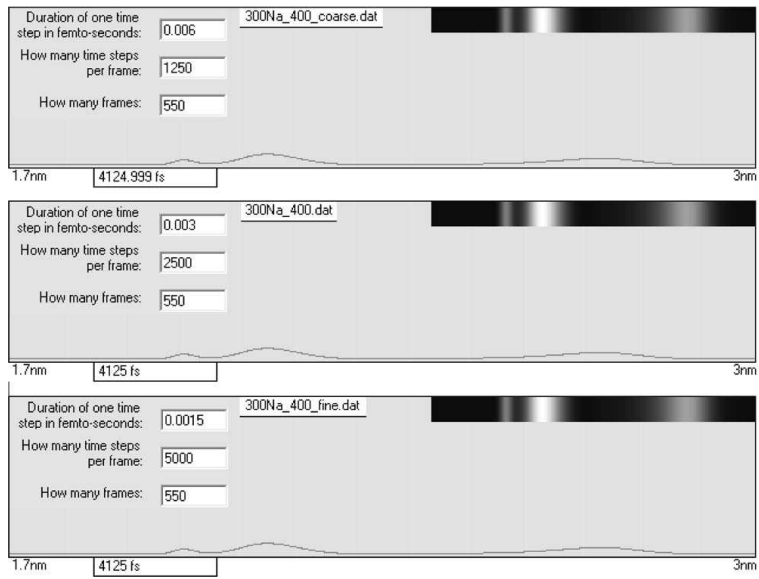}
\caption[Fig.4:]{Calculational consistency of the Crank-Nicolson approximation: a comparison of wave-function solutions with different temporal scales, ranging from 687.500 steps with a time width of 0.006 fs  (top)  to 1.375.000  steps (at 0.003 fs) (middle) and 2.750.000 steps (at 0.0015 fs) (bottom). Changes in the probability distribution of the ions spatial location (red curve at the bottom of the inserts) are not detectable, demonstrating the robustness and stabilitiy of the Crank-Nicolson method for solving equations of the type $ u_{t} = a.u_{xx}$ (e.g. the Schr\"{o}dinger equation) as used in the present study.}
\end{center}
\end{figure}

\section{Results}

\subsection{Conductive States}
In the following figures we provide some snapshots of varying potentials confining the quantum-mechanical ion $K_{Q}$ at location S1 of the filter region for different interactions with 'classical ions' KB and $K_{A}$ moving into the filter region. The filter sets out with a double occupancy of ions at location S1-S3 and develops into ion configuration S2-S4 with the outermost quantum-mechanically calculated ion close to S0. The snapshot shown in the following Figure 5 and Figure 6 cover a time-window of just 4.124 ps.

The time sequence of frames (a-d) in Figure 5 and their matching potentials (a-d) in Figure 6 cover the 'conductive' situation with one ion ($K_{B}$)  approaching the filter region from the cavity site, another ion ($K_{A}$) located at S3 and the wave-packet of the 'quantum-ion' ($K_{Q}$) centered at S1. During this time $(K_{Q})$ oscillates as a wave around it's center due to the thermal energy of the ion. In Figure 2 (b) the external ion exerts a repulsive 'knock-on' effect on the (K A). This external knock-on is also felt by the quantum-ion ( $K^{Q}$) as it now extends it's wave-function with some probability spreading into location S0. This non-local effect covers a z-extension in the order of $10^{-10}$m. Finally,  as seen in the frame Figure 5(d) the classical filter ion has reached position S2 inducing a strong non-local effect on the spread of the wave-function of the quantum ion, originally located at S1. There is now a considerable probability extension of 1nm (about the entire linear z-extension of the filter) for the amplitude of ion $K_{Q}$, with a spread into S0 and into the extra-cellular site of the filter.

\begin{figure}
\begin{center}
\epsffile{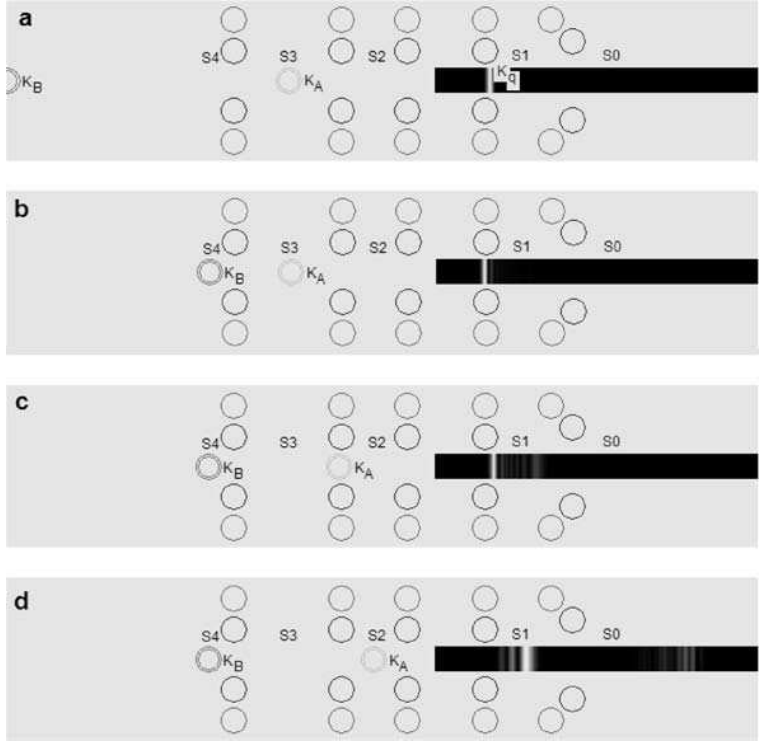}
\caption[Fig.5:]{QM-wave-packet distribution of one ion centered at location S1 interacting with two classical ions $K_{A}$ and $K_{B}$ that enter the filter region. The orientation of the filter atoms is from inside the cell (left) to outside the cell (right) according to the scheme shown in Figure 2(C). The wave-packet distribution of the ion $K_{Q}$ at location S1 is shown within dark stripes in order to visualize the emerging interference (coherence) pattern of its wavefunction. The confining potentials within these stripes are demonstrated in the following Figure 6 for four different time step situations, labeled by (a-d). The circles next to the ions, with two rows above and two rows below lining the filter, symbolize the location of C atoms (outside) and O atoms (inside) that belong to the carbonyl C=O groups of the filter. Note the distortion of C=O dipoles induced by the presence of the $K_{Q}$ ion in S1}
\end{center}
\end{figure}

\begin{figure}
\begin{center}
\epsffile{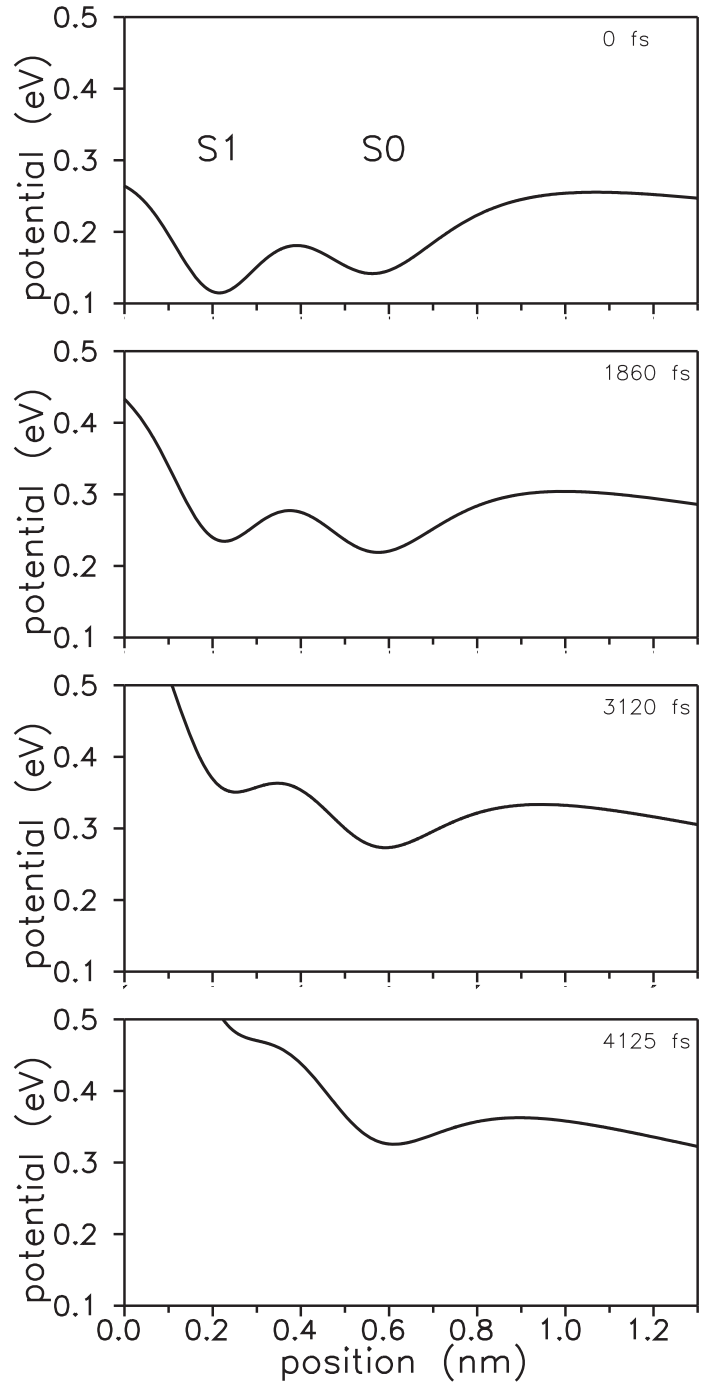}
\caption[Fig.6:]{Reconstructed dynamics of potentials along the filter locations S1 and S0 that results from the movement of 2 (classical) ions $K_{A}$, $K_{B}$ and one 'quantum-mechanical ' ion $K_{Q}$ during four consecutive snapshots (a-d) with locations as shown in Figure 2C.  The potentials are shown within the range of 0.1 to 0.5 eV. The x-axis is provided by the \textit{z}-coordinates of the filter region as shown in Figure 2B. It should be noted that the confining potentials for the ions are not pre-set, but the potential distribution emerges from the Coulomb interaction of ions with the charges provided by the lining oxygen atoms and the other ions present in the filter following the calculations by equations (2.2 - 2.4)}
\end{center}
\end{figure}

\subsection{Non-conductive states}

For the non-conductive states of the filter we assume that the classical ions occupy the cavity position and the filter position S3 initially. The quantum-ion remains in position S1 (the so-called S1-S3 filter configuration). We now allow the interacting C=O groups to oscillate thermally. This will in turn induce an oscillation of local potentials around the quantum-ion at S1.

\textbf{Cooling effects:}

It can be expected that the translocation of ions, followed by a transient binding into a local potential, will generally lead to a loss of energy of this ion and a subsequent '\textit{cooling effect}'. In an attempt at understanding the role of the filter structure for these possible effects, we have made a series of measurements on the dependence of mean kinetic energies on different oscillation frequencies of the coordinating C=O groups. The initial settings placed the ($K_{Q}$) ion into S1 with a kinetic energy of 20 meV (1 kT = 26 meV at room temperature),  the wave-packet extension was preset to 0.025 nm and the oscillatory amplitude of C=O to 0.02 nm. We followed the evolution of the $K^{Q}$ wave-function during a period of 9 ps at intervals of 0.006 fs (over 1.5 million steps).  In Figure 7 we show the mean kinetic energy of this ion at the end of the 9 ps cooling period as a function of the oscillation frequency of the C=O groups. The temporal development of the kinetic energies over the time window of 9 ps is demonstrated in Figure 8.

\begin{figure}
\begin{center}
\epsffile{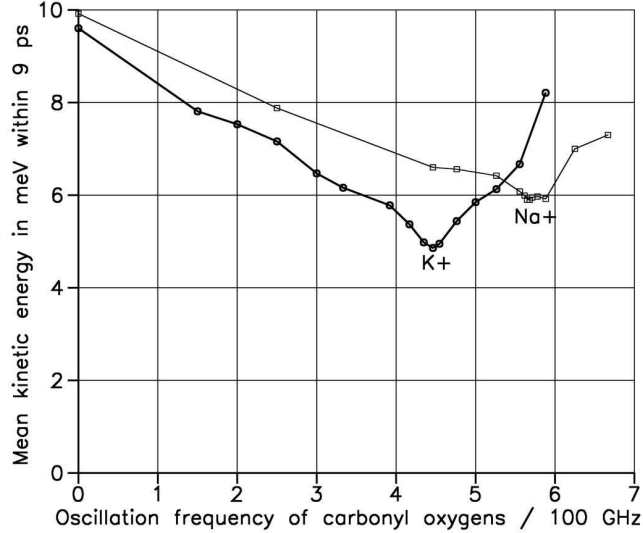}
\caption[Fig.7:]{Mean kinetic energy of a $K_Q^+i$ ion and a $Na_Q^+$ ion as a function of oscillation frequencies of filter C=O groups during 9 ps of simulation. The replacement of the 'intrinsic' $K_Q^+$ ion with a $Na_Q^+$ ion with different coordination distances to the surrounding atoms becomes reflected by a different kinetic energy minimum along the oscillation frequency of carbonyl dipoles. The loss of kinetic energy ('cooling') is less for the replaced atom species and requires a different oscillation frequency of C=O groups.}
\end{center}
\end{figure}

\begin{figure}
\begin{center}
\epsffile{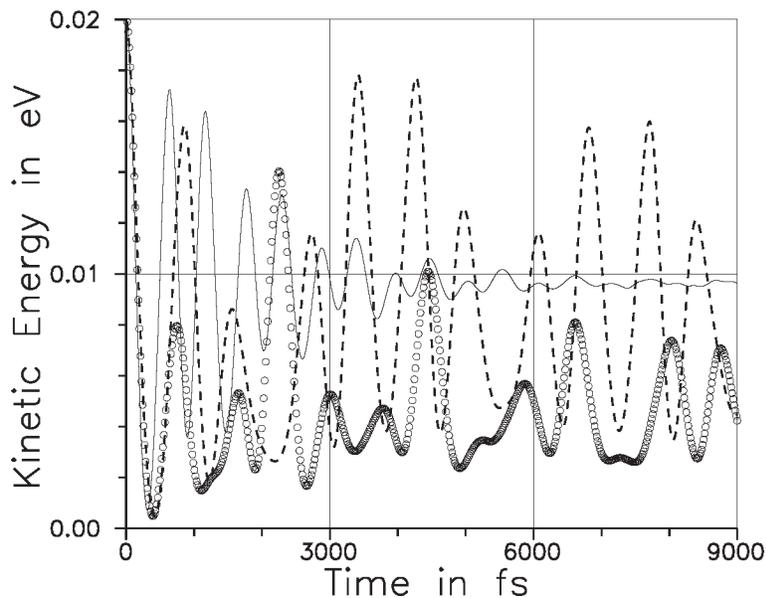}
\caption[Fig.8:]{The temporal evolution of kinetic energies of the $K_Q^+$ion as shown in Figure 7 at site S1 for three different oscillation frequencies: unbroken line = no oscillations, broken line = 588 GHz oscillations and circles = the 'optimal frequency' at 446 GHz with maximal cooling effects. The time evolution of these curves shows that only for the oscillation frequency that provides a clear cooling effect (Figure 7) at 446 GHz, the potential energy is not returned to the kinetic part during the next oscillation period. Thus the concerted quantum-mechanical behavior of the ion together with the coordinating carbonyl dipoles can account for this unique selection procedure at the atomic scale.}
\end{center}
\end{figure}

From Figure 8 it is apparent that at optimal oscillation frequencies of C=O groups (open circles in Fig.7) the kinetic energy approaches a minimum after a quarter cycle of the oscillation period. At this point the potential energy reaches a maximum. However, because the potential sink follows the movement of the ion, this potential energy is only partly returned into the kinetic term during the next quarter of the oscillation. This process continues until the kinetic energy of the ion approaches a minimum, depending on the frequency of the associated  C=O oscillations. Further, particular oscillation frequencies of the ion-carbonyl environment and the associated energy dissipation can discriminate different ion species for a given atomic filter environment. Whereas a primary ion selection effect seems to be associated by different ion-species dependent dehydration energies during the ions transfer from the cavity to the selectivity filter \cite{Re}, the present findings are strongly suggestive for an additional role of quantum oscillatory effects within the filters atomic environment on ion selectivity. This points to a clear quantum effect that is indispensible in explaining ion selectivity, a basic and indispensible biological function of ion channel proteins.
As expected, our results show that if an $Na^{+}$ ion is placed in the same position as a $K^{+}$ , it also can transfer energy to the C=O lattice (see Figure 7). However, since the $Na^{+}$ ion has a lower mass than the $K^{+}$ ion, the most efficient cooling occurs at a higher oscillation frequency of the C=O groups.  The natural oscillation frequency of the C=O groups caused by thermal motion of the surrounding atoms is essentially determined by the binding strength of the C=O groups to the back- bone of the protein strands. We can therefore expect this oscillation to exert its cooling effect in the most efficient way for ions with a specific charge and a specific mass. When hopping from one site to a neighboring site in the selectivity filter only a particular ion species will be captured optimally by the coordinating C=O cage. However, because the exchange of energy works both ways, the con-specific ion will also be able to extract energy from the coordinating C=O group within a shorter time than any other ion species. This in turn will allow shorter times for bridging the energy barrier to the neighboring site. Further simulations will be required to test whether this effect can lead to a possible resonance phenomenon accompanying a fast passage through the selectivity filter.

\section{Conclusions}
We have shown that, depending on the surrounding carbonyl derived potentials, alkali ions in the filter region of ion channels lined by the conserved TVGYG amino acid sequence can become highly 'quantum-delocalized' depending on ion location and oscillatory motion of the coordinating C=O groups. Further, we find that due to the interaction of ions with a time-dependent potential, the energy of ions is not conserved. Depending on the frequencies of oscillating C=O surroundings the ion can lose up to 1/2 of its kinetic energy, exerting a substantial 'cooling effect'. The transfer of this energy to the C=O environment is shown to depend on a specific C=O resonant oscillation frequency and within the present study is observed under a 'non-conductive' filter state for an ion at location S1. 
	In the frame of previous work on energy transfer systems in proteins, quantum and classical \cite{Br1,Sal,Sal2}we suggest that the observed effect will cool down the ions inside the filter giving off their energy to the environment, i.e. to the so-called P-loop domain of the lining filter region. Energy transfer along the P-loop backbone atoms can in turn change the symmetry of the filter entrance \cite{B3,Cord,Kona}. As a result it was anticipated that even the location of a single ion in the filter can become associated with the gating state of the selectivity filter \cite{B3,Cord}. Here we provide evidence for this situation based on a quantum-mechanical calculation and offer predictions about the involved energies and resonant oscillations and their effect on channel filter states. These results indicate that temporal development of quantum states in channel proteins can propagate into classical ion-channel conformations that determine the electrical signal properties of neuronal membranes. Because the filter states must be correlated with the pore-domain gating state of the channel in order to facilitate the co-occurrence of conduction and access of ions to the channel protein \cite{Van}, the pore gating states of the channels can be interpreted as 'classical witness  states' \cite{Ved} of an underlying quantum process in the brain, as suggested previously \cite{GB1,GB2,Roy}.



\end{document}